\documentclass[
preprintnumbers,
nofootinbib,
 amsmath,amssymb,
 prd,aps,superscriptaddress,twocolumn,
floatfix,
]{revtex4-2}

\usepackage{aas_macros}
\usepackage{graphicx}
\usepackage{amsmath}
\usepackage[caption=false]{subfig}
\usepackage{siunitx}
\usepackage{placeins}
\usepackage{color}
\usepackage{standalone}
\usepackage{dcolumn}
\usepackage{tensor}
\usepackage{bm}
\usepackage{microtype}
\usepackage{etoolbox}
\usepackage{amssymb}
\usepackage{mathrsfs}
\usepackage{accents}
\usepackage[normalem]{ulem}
\usepackage{soul} % For \st{} and \hl{}
\usepackage[dvipsnames]{xcolor}
\usepackage[colorlinks,urlcolor=NavyBlue,citecolor=NavyBlue,linkcolor=NavyBlue,pdfusetitle]{hyperref}
\usepackage[all]{hypcap}
\usepackage[inline]{enumitem}
\usepackage[utf8]{inputenc}
\usepackage{xspace}
\usepackage[printonlyused, nolist]{acronym}
\usepackage{lineno}
\graphicspath{{./figures/}}

\newcommand{\msun}{\ensuremath{M_{\odot}}\xspace}
\newcommand{\fPBH}{\ensuremath{f_{\rm PBH}}\xspace}
\newcommand{\aIII}{\ensuremath{a_{\rm III}}\xspace}
\newcommand{\bIII}{\ensuremath{b_{\rm III}}\xspace}
\newcommand{\zIII}{\ensuremath{z_{\rm III}}\xspace}
\newcommand{\nIII}{\ensuremath{\dot{n}_{\rm III}}\xspace}
\newcommand{\pIII}{\ensuremath{p_{\rm III}}\xspace}
\newcommand{\betaPBH}{\ensuremath{\beta_{\rm PBH}}\xspace}
\newcommand{\pPBH}{\ensuremath{p_{\rm PBH}}\xspace}
\newcommand{\npbh}{\ensuremath{\dot{n}_{\rm PBH}}\xspace}
\newcommand{\RateDensityUnit}{\ensuremath{~\mathrm{{Gpc}^{-3}{yr}^{-1}}}\xspace}
\newcommand{\Nobs}{\ensuremath{N_{\rm obs}}}
\newcommand{\dNdz}{\ensuremath{\frac{dR}{dz}}\xspace}

\newcommand{\lp}{\left (}
\newcommand{\rp}{\right )}
\newcommand{\llp}{\left [}
\newcommand{\rrp}{\right ]}
\renewcommand{\d}{{\rm d}}
\newcommand{\PBH}{{\rm PBH}}

\newcommand{\jhu}{\affiliation{Department of Physics and Astronomy, Johns Hopkins University, 3400 N. Charles
Street, Baltimore, MD 21218, USA}}

\newcommand{\ligoMIT}{\affiliation{LIGO Laboratory, Massachusetts Institute of Technology, Cambridge, Massachusetts 02139, USA}}
\newcommand{\MKIMIT}{\affiliation{Kavli Institute for Astrophysics and Space Research and Department of Physics, Massachusetts Institute of Technology, Cambridge, Massachusetts 02139, USA}}
\newcommand{\geneva}{\affiliation{D\'epartement de Physique Th\'eorique and Centre for Astroparticle Physics (CAP), Universit\'e de Gen\`eve, 24 quai E. Ansermet, CH-1211 Geneva, Switzerland}}
\newcommand{\sapienza}{\affiliation{Dipartimento di Fisica, Sapienza Università 
di Roma, Piazzale Aldo Moro 5, 00185, Roma, Italy}}
\newcommand{\infn}{\affiliation{INFN, Sezione di Roma, Piazzale Aldo Moro 2, 00185, Roma, Italy}}

\begin{document}

%\preprint{ET-0068A-22}

\title{Constraining high-redshift stellar-mass primordial black holes\\with next-generation ground-based gravitational-wave detectors}
%\title{Constraining stellar-mass primordial black hole abundance with the next-generation ground-based gravitational-wave detectors}
\author{Ken~K.~Y.~Ng}
\email{kenkyng@mit.edu}
\ligoMIT
\MKIMIT

\author{Gabriele~Franciolini}
\email{gabriele.franciolini@uniroma1.it}
\sapienza
\infn
\author{Emanuele~Berti} 
\jhu

\author{Paolo~Pani}
\sapienza
\infn

\author{Antonio~Riotto}
\geneva

\author{Salvatore~Vitale}
\ligoMIT
\MKIMIT

\date{\today}
\begin{abstract}
The possible existence of primordial black holes in the stellar mass window has received considerable attention because their mergers may contribute to current and future gravitational-wave detections.
Primordial black hole mergers, together with mergers of black holes originating from Population~III stars, are expected to dominate at high redshifts ($z\gtrsim 10$).
However the primordial black hole merger rate density is expected to rise monotonically with redshift, while  Population~III mergers can only occur after the birth of the first stars.
Next-generation gravitational-wave detectors such as Cosmic Explorer~(CE) and Einstein Telescope~(ET) can access this distinctive feature in the merger rates as functions of redshift, allowing for a direct measurement of the abundance of the two populations, and hence for robust constraints on the abundance of primordial black holes. We simulate four-months worth of data observed by a CE-ET detector network and perform hierarchical Bayesian analysis to recover the merger rate densities. We find that if the Universe has no primordial black holes with masses of $\mathcal{O}(10\msun)$, the projected upper limit on their abundance $\fPBH$ as a fraction of dark matter energy density may be as low as $\fPBH\sim \mathcal{O}({10^{-5}})$, about two orders of magnitude lower than current upper limits in this mass range. If instead $\fPBH\gtrsim 10^{-4}$, future gravitational wave observations would exclude $\fPBH=0$ at the 95\% credible interval.
\end{abstract}

\maketitle

\section{Introduction}
Over the past few decades, it was suggested that primordial black holes (PBHs) may be formed in the early Universe, shortly after the Big Bang~\citep{pbh1966zeldovich,Carr:1974nx,Hawking:1971ei}.
Although various experiments constrain the abundance of these objects (see e.g.~\cite{Carr:2020gox} for a recent review), in certain mass ranges PBHs could comprise a significant fraction of the dark matter, seed supermassive BHs at high redshift~\cite{2010A&ARv..18..279V,Clesse:2015wea,Serpico:2020ehh}, and contribute to current~\cite{Clesse:2020ghq,DeLuca:2020sae,Franciolini:2021tla} and future~\cite{DeLuca:2021wjr, DeLuca:2021hde, Pujolas:2021yaw,Auclair:2022lcg} gravitational-wave (GW) observations.

PBHs generated from the collapse of sizable Gaussian overdensities in the radiation-dominated early Universe~\cite{Ivanov:1994pa,GarciaBellido:1996qt,Ivanov:1997ia,Blinnikov:2016bxu} are expected to possess small spins~\cite{DeLuca:2019buf, Mirbabayi:2019uph}, and are not clustered~\cite{Ali-Haimoud:2018dau,Desjacques:2018wuu,Ballesteros:2018swv,MoradinezhadDizgah:2019wjf,Inman:2019wvr,DeLuca:2020jug} at the formation epoch.
In this scenario, they may gravitationally decouple from the Hubble flow before the matter-radiation equality and get bound in binaries~\cite{Nakamura:1997sm,Ioka:1998nz} (see~\cite{Sasaki:2018dmp,Green:2020jor,Franciolini:2021nvv} for reviews).
These high-redshift (i.e. $z\gtrsim 10^3$) PBHs can produce merging binary black holes (BBHs) of primordial origin, with a merger rate that is expected to increase monotonically with redshift.

High-redshift astrophysical BBHs may also originate from Population~III (Pop~III) stars~\citep{Schneider:1999us,Schneider:2001bu,Schneider:2003em,Kinugawa:2014zha,Kinugawa:2015nla,Hartwig:2016nde,Belczynski:2016ieo}.
Since these BBHs can only form after the collapse of Pop~III stars in the matter-dominated era, the Pop~III BBH merger rate is expected to decay at high redshifts, as we approach the birth epoch of the first stars.
While observations remain elusive, estimates from cosmological theories and simulations suggest that the first stars might have been born at redshifts $z\lesssim 50$~\citep{Bromm:2005ep,Tornatore:2007ds,Trenti:2009cj,deSouza:2011ea,Koushiappas:2017kqm,Mocz:2019uyd}.
Together with the typical time delay between formation and merger for Pop~III stars, $\mathcal{O}(10)~\rm Myr$, one expects the epoch of Pop~III mergers to start at $z\lesssim 40$ and peak at $z\lesssim 10$~\citep{Kinugawa:2014zha,Kinugawa:2015nla,Hartwig:2016nde,Belczynski:2016ieo,Inayoshi:2017mrs,Liu:2020lmi,Liu:2020ufc,Kinugawa:2020ego,Tanikawa:2020cca,Tanikawa:2021qqi}.

Observations of these high-redshift populations will be made possible by the proposed next-generation (NG) ground-based GW detectors~\cite{Kalogera:2021bya}, Cosmic Explorer (CE)~\citep{Evans:2016mbw,Reitze:2019iox,CEHS} and Einstein Telescope (ET)~\citep{Punturo:2010zz,Maggiore:2019uih}, both of which will be able to observe BBHs with total mass of $\mathcal{O}(10-100)~\msun$ at redshifts up to $z\sim 100$~\cite{Hall:2019xmm}.
Combining these high-redshift BBH observations allows for precise measurements of the branching ratios and detailed properties of PBH and Pop~III mergers, and thus for robust constraints on the PBH abundance (typically presented in terms of the fraction $\fPBH\equiv \Omega_{\rm PBH}/\Omega_{\rm DM}$ of the dark matter energy density in the form of PBHs) from direct measurements in the $\sim [10-50]~\msun$ mass window, which is the most sensitive mass range of CE and ET~\cite{Evans:2021gyd}.

In this study we only make use of a simple, general property of the PBH and Pop~III merger rate densities: the former rises monotonically, while the latter decays rapidly as the redshift increases.
We first review theoretical predictions for the merger rate densities of these two high-redshift populations and the Bayesian statistical framework in the population analysis.
We then simulate four-months worth of data in a detector network of CE and ET with realistic redshift uncertainty estimates, and infer the morphology of the merger rate densities at $z\geq8$ relying on redshift measurements only.
In the absence of PBH mergers, the derived upper limit on $\fPBH$ is almost two orders of magnitude lower than current constraints.
If a fraction of PBHs as low as $\fPBH\sim 10^{-4}$ exists in the Universe, we show that $\fPBH$ could be measured precisely. 
Finally we discuss why our results are conservative, and how one could further improve the constraints by including any other feature that might help distinguishing the two channels, namely BBH mass, spin and eccentricity. 

\section{Merger rates densities at high redshift}
\subsection{Primordial Black Hole mergers}
Throughout this paper, we assume the standard PBH formation scenario describing the collapse of large Gaussian overdensities in the radiation dominated early Universe 
\cite{Zeldovich:1967lct,Hawking:1974rv,Chapline:1975ojl,Carr:1975qj,
Ivanov:1994pa,GarciaBellido:1996qt,Ivanov:1997ia,Blinnikov:2016bxu,Musco:2020jjb,Escriva:2021aeh}.
In this setting, PBHs follow a Poisson spatial distribution at formation~\cite{Ali-Haimoud:2018dau,Desjacques:2018wuu,Ballesteros:2018swv,MoradinezhadDizgah:2019wjf,Inman:2019wvr},
which triggers PBH binary formation already at very high redshift. Indeed,  
it was shown that this process typically takes place before the matter-radiation equality and it is 
due to gravitational attraction leading to PBH pairs decoupling from the Hubble flow~\cite{Nakamura:1997sm,Ioka:1998nz}. We will assume that the PBH mass distribution is described by a lognormal function
\begin{equation}
    \psi(m | M_c, \sigma ) = 
    \frac{\exp 
    \llp 
    -{\log ^2 (m/M_c)}/{2 \sigma^2} 
    \rrp
    }{\sqrt{2\pi}\sigma  m}\,,
\label{eq:masspsi}
\end{equation}
centered at the mass scale $M_c$ (not to be confused with the binary chirp mass) with width $\sigma$. This model-independent parametrization of the mass function can describe
a population arising from a symmetric peak in the power spectrum of curvature perturbations in a wide variety of formation models (see e.g. Refs.~\cite{Dolgov:1992pu,Carr:2017jsz}) and is often used in the literature to
set constraints on the PBH abundance from GW measurements~\cite{Garcia-Bellido:2017fdg,Raidal:2018bbj,DeLuca:2020qqa,DeLuca:2020sae,Wong:2020yig,Gow:2019pok,Hall:2020daa,Bhagwat:2020bzh,Hutsi:2020sol,DeLuca:2021wjr,Franciolini:2021tla,Bavera:2021wmw}. 

PBHs are subject to accretion of baryonic matter during their cosmic evolution, which can impact both their mass and spin distributions~\cite{DeLuca:2020bjf,DeLuca:2020fpg}, as well as hardening PBH binaries and enhancing the merger rate~\cite{DeLuca:2020qqa}. We refer to Ref.~\cite{DeLuca:2020qqa} for a thorough discussion of uncertainties in the accretion model. 
In the mass range of interest for next-generation detectors, similar to the one currently probed by the LIGO/Virgo/KAGRA Collaboration (LVKC) detectors,
accretion is never efficient for $z\gtrsim 30$, while it may have an impact on the PBH population at smaller redshift. 
In this work, we neglect the role of accretion. This means that the constraints we set on the PBH abundance based on merger rate estimates are the {\it least} stringent, and may be translated to smaller values of $f_\PBH$ if a strong accretion phase is assumed. This is because accretion enhances the merger rate, allowing to probe smaller abundances. 

We compute the differential volumetric PBH merger rate density following~Refs.~\cite{Raidal:2018bbj, Vaskonen:2019jpv,DeLuca:2020jug,DeLuca:2020qqa} as
\begin{align}
\label{eq:diffaccrate}
 \frac{\d  \npbh}{\d m_1 \d m_2}
& = 
\frac{1.6 \times 10^6}{{\rm Gpc^3 \, yr}} 
f_\PBH^{\frac{53}{37}} 
\lp \frac{t(z)}{t_0} \rp^{-\frac{34}{37}}  
\eta^{-\frac{34}{37}}
\lp \frac{M}{M_\odot} \rp^{-\frac{32}{37}}  
 \nonumber \\
& \times
S(M, f_\PBH)
\psi(m_1) \psi (m_2)
\end{align}
where 
 $M = m_1+m_2$,  $\eta = m_1 m_2/M$, and $t_0$ is the current age of the Universe. The suppression factor $S<1$  accounts for environmental effects in both the early- and late-time Universe.
In the early Universe, these are a consequence of interactions between PBH binaries and both the surrounding dark matter inhomogeneities, as well as neighboring PBHs at high redshift~\cite{Eroshenko:2016hmn,Ali-Haimoud:2017rtz,Raidal:2018bbj,Liu:2018ess}. 
In the late Universe, they are due to successive disruption of binaries which populate PBH clusters formed from the initial Poisson conditions~\cite{Vaskonen:2019jpv,Jedamzik:2020ypm,Young:2020scc,Jedamzik:2020omx,DeLuca:2020jug,Trashorras:2020mwn,Tkachev:2020uin,Hutsi:2020sol,link} throughout the evolution of the Universe.
An analytic expression for $S$ can be found in Ref.~\cite{Hutsi:2020sol} and, for the small abundance values of interest here, it can be approximated as 
\begin{align}
    &S(M, f_\PBH) \simeq 33 f_\PBH^{21/37} 
    + {\cal O}\lp \frac{f_\PBH}{10^{-3}}\rp^{58/37}\,.
\end{align}
Therefore, assuming a small enough abundance $f_\text{\tiny PBH} \ll  10^{-3}$, one can write the PBH \textit{merger mass function} as
\begin{align}\label{eq:pbbhMassFunction}
    p(m_1, m_2) \propto
    {(m_1+m_2)^{-\frac{36}{37}}}{(m_1 m_2)^{-32/37}}
\psi(m_1) \psi (m_2)
\end{align}

This reveals a key feature that we will exploit in our analyses, namely that the volumetric merger rate density of PBHs has a power-law dependence on the age of the Universe $t(z)$ extending up to $z\gtrsim 10^3$:
\begin{align}\label{eq:pbhdRdz}
    \npbh(z) \propto \left(\frac{t(z)}{t_0}\right)^{-34/37}.
\end{align}
Given a choice of mass function and $\fPBH$, the local rate density of PBH mergers can be obtained by integrating Eq.~\eqref{eq:diffaccrate}, evaluated at redshift $z=0$, over $m_1$ and $m_2$. 
The result of this integration will be used below to translate posterior distributions of the local PBH merger rate to constraints on the abundance $f_\PBH$.

\subsection{Pop~III mergers}\label{sec:popIIIrate}
We use the phenomenological model for the volumetric merger rate density of Pop~III BBHs from Ref.~\cite{Ng:2020qpk}. This model is a simple fit to the merger rate density predicted from population synthesis studies~\cite{Belczynski:2016ieo}:
\begin{align}\label{eq:pop3dRdz}
\nIII(z \mid \aIII, \bIII, \zIII) \propto \frac{e^{\aIII(z-\zIII)}}{\aIII+\bIII e^{(\aIII+\bIII)(z-\zIII)}},
\end{align}
where $\aIII$, $\bIII$, and $\zIII$ characterize the upward slope at~$z<\zIII$, the downward slope at~$z>\zIII$, and the peak location of the volumetric merger rate density, respectively.
The normalization of $\nIII$ is the same as in Ref.~\citep{Ng:2020qpk}: we first choose the ratio between the peak of $\nIII$ and the Pop~I/II merger rate density to be $1/10$ (supported by population synthesis studies, e.g.~\citep{Belczynski:2016ieo,Tanikawa:2021qqi}), and obtain $\nIII(\zIII)=20\RateDensityUnit$ by matching the simulated Pop~I/II merger rate density~\citep{Belczynski:2016obo,Ng:2020qpk} with a local merger rate density of $25\RateDensityUnit$, consistent with current measurements~\citep{GWTC3,GWTC3rate}.

The initial conditions of Pop~III stars, and hence the mass spectrum of Pop~III BBHs, are highly uncertain~\citep{Belczynski:2016ieo,Tanikawa:2020cca,Tanikawa:2020abs,Kinugawa:2020tbg,Kinugawa:2020ego,Kinugawa:2020xws,Inayoshi:2017mrs,Tanikawa:2021qqi,Hijikawa:2021hrf}.
We further assume the mass spectrum of Pop~III BBHs to be exactly the same as that of PBH mergers.
This implies that the mass spectra of the two populations are indistinguishable, and only redshift measurements are informative in the inference of branching ratios between the two populations. This is a conservative approach, as differences in the mass spectrum could be used to distinguish between the two populations, yielding more precise measurements than the ones we present below.

\section{Statistical model}
It is more convenient to work with the normalized redshift distribution of the $k$-th subpopulation,
\begin{align}
    p_{k}(z) &\propto \frac{\dot{n}_k(z)}{1+z} \frac{dV_c}{dz},
\end{align}
where $dV_c/dz$ is the differential comoving volume and the factor of $1+z$ accounts for the cosmological time dilation.
The differential merger rate of the whole population in the detector frame as a function of redshift is then
\begin{align}\label{eq:dRdz}
    \frac{dR}{dz} &= R\left[\betaPBH \pPBH(z) + (1-\betaPBH) \pIII(z)\right],
\end{align}
where $\betaPBH$ is the fraction of the PBH merger rate to the total merger rate, and $R=\int (dR/dz) dz$ is the total merger rate in the detector frame.
In this phenomenological model, we have five hyperparameters: $(\aIII, \bIII, \zIII, \betaPBH, R)$.

Given $\Nobs$ BBH detections, $\pmb{d}=\{d_i\}^{\Nobs}_{i=1}$, we can use hierarchical Bayesian inference to extract the posterior of the hyperparameters of $dR/dz$~\cite{Farr:2013yna,Mandel:2018mve,Thrane:2018qnx,Wysocki:2018mpo,Vitale:2020aaz}.
Since we focus on the redshift range in which the GW detectors are not fully sensitive to the emission from BBHs~\cite{Hall:2019xmm}, we have to take into account the selection effect due to the detection efficiency as a function of redshift~\cite{Mandel:2018mve,Thrane:2018qnx,Wysocki:2018mpo,Vitale:2020aaz}.
In this study we only infer $dR/dz$, but not the mass function of the two populations.
To achieve this, we assume that the BBH masses measurements of individual events are completely uninformative.
The mathematical details of the statistical inference are given in Appendix~\ref{app:hba}.

The phenomenological hyperparameters implicitly depend on the physical parameter of interest, i.e. the PBH abundance $\fPBH$. To obtain the posterior of $\fPBH$, one first obtains the posterior of the PBH merger rate density by subtracting the contribution of Pop~III mergers from Eq.~\eqref{eq:dRdz}. {Then, if the two PBH mass function parameters, $(M_c, \sigma)$ are known or measured, one can convert the posterior of $\npbh(0)$ to the posterior of $\fPBH$ through the normalization in Eq.~\eqref{eq:diffaccrate}. }
Besides the phenomenological model discussed above, we also employ Gaussian process regression (GPR) to repeat the analysis with minimal assumptions.
In this unmodeled approach, we parameterize $dR/dz$ as a piecewise-constant function over several redshift bins, and assume that the rate of each bin is drawn from a Gaussian process with a squared-exponential kernel.
This model does not contain any information on the two subpopulations, hence one cannot directly infer $\fPBH$ without further modeling assumptions.
However, it serves as an independent cross-check of the phenomenological approach in which model systematics may be of concern.
The mathematical details of GPR are given in Appendix~\ref{app:GPR}.

\section{Simulations}\label{Sec.Sims}
In the following, we consider five simulated universes with different ratios of $\betaPBH=0$, 1/9, 1/5, 1/3, and 1/2.
The GW detector network consists of one CE in the United States and one ET in Europe.
The Pop~III merger rate density is fixed to the benchmark values of $( a_{\rm III}, b_{\rm III}, z_{\rm III} ) = (0.66, 0.3, 11.6)$, which are the best-fit values\footnote{While the model of Eq.~\eqref{eq:pop3dRdz} with these values is a good fit to the simulated rate in Ref.~\cite{Belczynski:2016ieo} around the peak $(8\lesssim z\lesssim15)$, we found that it may overestimate both lower and higher redshift tails by a factor of a few.
This introduces a higher noise floor for mapping the redshift ``tail'' of PBH mergers at $z\gtrsim 30$, and hence leads to a more conservative estimate of the upper limit in $\fPBH$.}
 to the simulations in Ref.~\cite{Belczynski:2016ieo}.
We simulate 800 Pop~III BBHs and $800~\betaPBH/(1-\betaPBH)$ primordial BBHs whose true redshifts are drawn from $\nIII$ and $\npbh$, respectively, and true masses are sampled from the same mass function in Eq.~\eqref{eq:pbbhMassFunction} with $(M_c, \sigma)=(30 \msun, 0.3)$, which encompasses the most sensitive mass range of the next-generation GW detectors.
This amounts to four-months worth of data.
To create a mock data set of redshift measurements, we assume a lognormal likelihood for each simulated event.
The width of the likelihood scales inversely with the signal-to-noise ratio (SNR) of the signal, and is calibrated by the simulated catalog of the full Bayesian posteriors using \textsc{IMRPhenomPv2} waveforms~\citep{Hannam:2013oca,Husa:2015iqa,Khan:2015jqa,Vitale:2016icu} (see Sec.~\ref{Sec.Dis} for a discussion about higher-order modes).
The details of the calibration are presented in Appendix~\ref{app:mockLike}.

As we are only interested in the high-redshift region, where the astrophysical BBHs originated from Pop~I/II stars are negligible, we downselect the events whose true redshifts are above 8.
This is justified by the comparison between the Pop~I/II merger rates and the Pop~III merger rate in simulation studies~\citep{Belczynski:2016ieo,Tanikawa:2021qqi}.
Finally, we discard signals which are below a network SNR threshold of 12.
The remaining events form the total set of $\Nobs$ detections.
We note that $\Nobs$ depends on the merger mass function and total merger rate, since the SNR depends on both the masses and redshifts of the BBHs.

\section{Results}

\begin{figure}[t]
    \centering
    \includegraphics[width=0.95\columnwidth]{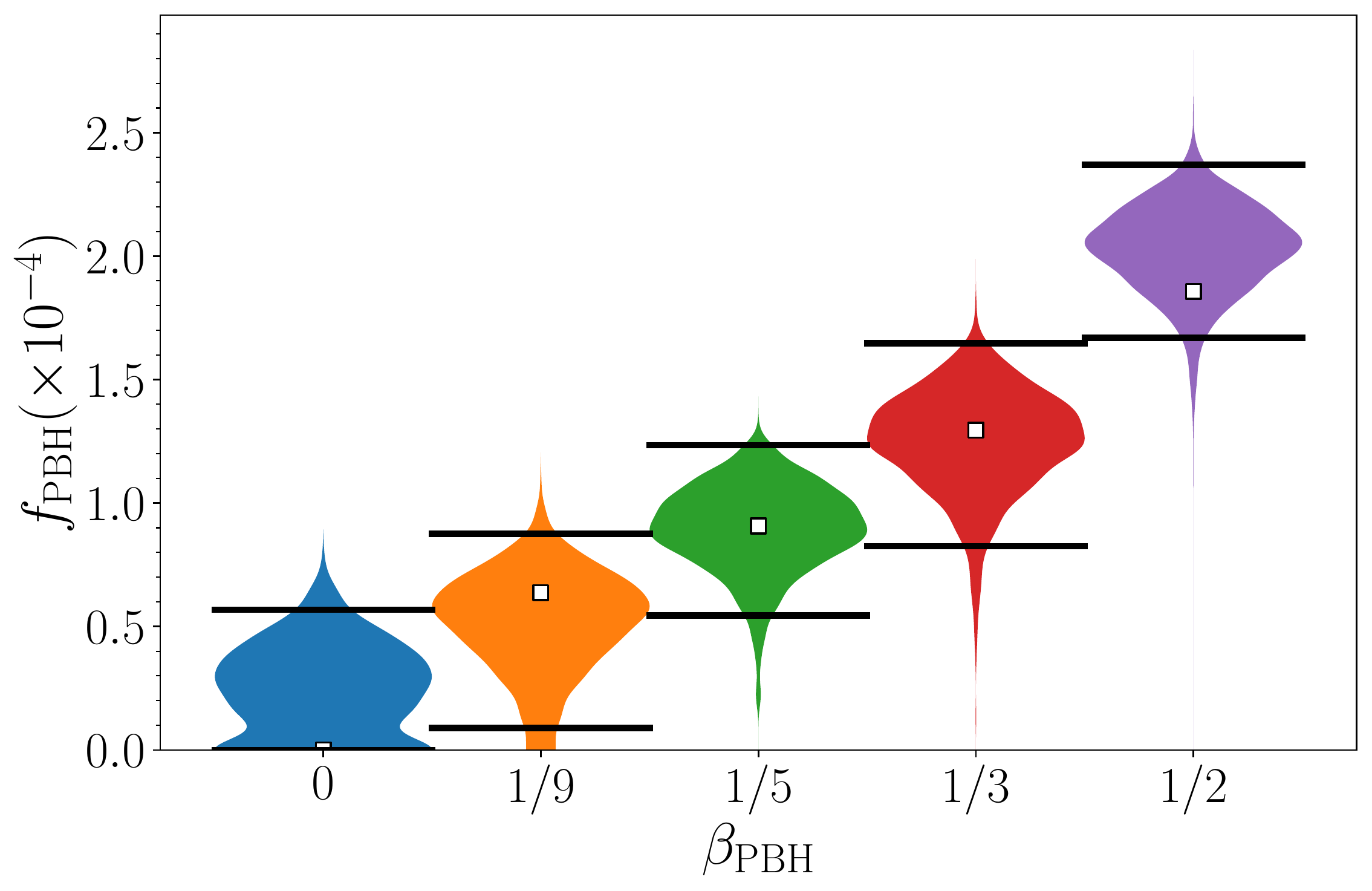}
    \caption{Posteriors of $\fPBH$ given $(M_c, \sigma)=(30 \msun, 0.3)$ at $\betaPBH=0,1/9, 1/5, 1/3,$ and $1/2$.
    The true values of $\fPBH$ corresponding to $\betaPBH$ are marked by squares in each violin plot.
    The solid black lines indicate the 95\% CIs.
    The prior is reweighted to be uniform in $\fPBH$.
    }
    \label{fig:fPBHPosterior}
\end{figure}

We first compare the measurement of $\fPBH$ in all universes derived from the modeled hierarchical inference.
The resulting posteriors of $\fPBH$,
\begin{align}
    p(\fPBH \mid M_c=30\msun,~\sigma=0.3,~\pmb{d}),\nonumber
\end{align}
in the various universes we simulated are shown in Fig.~\ref{fig:fPBHPosterior}. 
{This expression assumes that $M_c$ and $\sigma$ are known and equal to the ``true'' values used to generate the simulated signals, while in reality one should marginalize over those unknown parameters. However, we observe that varying $M_c$ within $[10, 50]~\msun$, which is the typical uncertainty in the source-frame primary mass at $z\sim 30$~\cite{Ng:2021sqn}, only changes the true value of $\fPBH$ by $\mathcal{O}(10\%)$.
Similarly, the variation of $\sigma$ in $[0.1, 0.5]$ has a negligible impact on the values of $\fPBH$.
As such, we skip the marginalization over the uninformative prior of $(M_c, \sigma)$, which would roughly average out the variation in the values of $\fPBH$ within the mass range of interest. }
Even with four months of observations, the posteriors for $\betaPBH\geq 1/9 $ exclude  $\fPBH=0$ within the 95\% highest-posterior-density credible interval (CI).
This implies we can \textit{measure} $\fPBH\sim10^{-4}$ with a relative uncertainty of $\lesssim 50\%$ in realizations of a Universe with more than 10\% PBH mergers compared to Pop~III mergers.
In a Universe without PBH mergers $(\betaPBH=0)$, we can \textit{constrain} the upper limit of $\fPBH\lesssim 6\times 10^{-5}$ at the 95\% CI, leading to more than an order-of-magnitude improvement when compared to the existing constraints in this mass range, which are dominated by the one drawn from the current GW LVKC data.

\begin{figure}[!h]
    \centering
    \includegraphics[width=0.98\columnwidth]{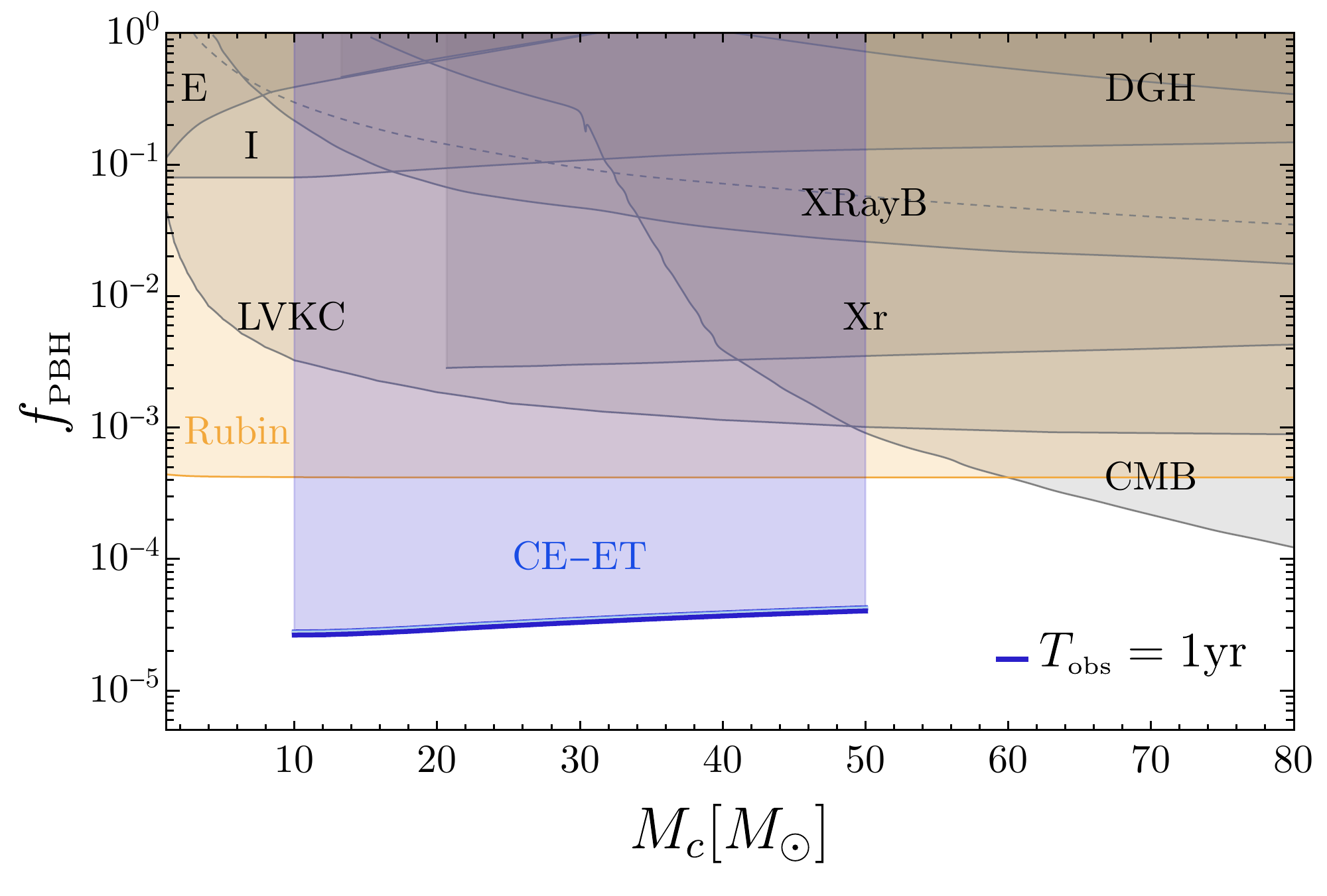}
    \caption{Projected upper limit on $\fPBH$ as a function of $M_c$ drawn from the merger rate density measurement in the simulated Universe without PBHs, and scaled to one year of observations.We also show the current most stringent constraints in this mass range coming from microlensing searches (assuming Poisson inital condition at formation~\cite{Petac:2022rio,Gorton:2022fyb}) by MACHO/EROS (E)~\cite{Alcock:2000kd, Allsman:2000kg} and  Icarus (I)~\cite{Oguri:2017ock};  measurements of galactic X-rays (Xr)~\cite{Manshanden:2018tze} and X-Ray binaries (XRayB)~\cite{Inoue:2017csr}; CMB distortions by spherical or disk accretion (Planck S and Planck D, respectively)~\cite{Ali-Haimoud:2016mbv, Serpico:2020ehh}; and Dwarf Galaxy heating (DGH)~\cite{Lu:2020bmd,Takhistov:2021aqx}. LVKC stands for the constraint coming from the LIGO/Virgo/Kagra Collaboration merger rate measurements \cite{Ali-Haimoud:2017rtz,Raidal:2017mfl,Vaskonen:2019jpv,Wong:2020yig,Franciolini:2021tla}. We neglect the role of accretion, which has been shown to affect constraints on masses larger than ${\cal O}(10)~M_\odot$~\cite{DeLuca:2020fpg,DeLuca:2020qqa,DeLuca:2020bjf}. For more details, see the review in Ref.~\cite{Carr:2020gox}. In yellow, we show forecasts for the limits that will be set by microlensing searches with the Rubin observatory~\cite{LSSTDarkMatterGroup:2019mwo,2022arXiv220308967B}.}
\label{fig:fPBHlimit}
\end{figure}
Next, we explore how differences in the PBH mass spectrum affect the projected upper limit.
To bypass the need of repeating the mock data analysis for each set of true $(M_c, \sigma)$, we account for the relative change of $\Nobs$ caused by the variation of the merger mass function.
Making the reasonable assumption that the statistical uncertainty in the hierarchical inference has reached the $1/\sqrt{\Nobs}$ regime, then our goal is accomplished by scaling the uncertainty of $\fPBH$ with $\sqrt{\Nobs(M_c=30\msun)/\Nobs(M_c)}$.

In Fig.~\ref{fig:fPBHlimit}, we show the projected upper limit of $\fPBH$ as a function of $M_c\in[10, 50]~\msun$ with one year\footnote{This result is obtained from our 4 months worth of simulations by scaling the uncertainties as $T_{\rm{obs}}^{-1}$.} of observations (blue lines, nearly entirely overlapping), and compare it with the current constraints (see the caption). {Different blue curves correspond to different values of $\sigma$ in the range $[0.1-0.5]$.}
This constraint worsens quickly for larger or smaller masses as the NG detectors' horizon shrinks and the number of observations at high redshift drastically decreases outside of the mass range we consider.
 Fig.~\ref{fig:fPBHlimit} shows how NG detectors can improve the upper limit derived from the current GW observations (LVKC curve) within the stellar-mass window by almost two orders of magnitude.
In the near future, the number of BBH observations made by the advanced detectors at design sensitivity may increase by a factor of $\sim 10$ within $z\lesssim 2$~\citep{TheLIGOScientific:2014jea,TheVirgo:2014hva,KAGRA:2020tym,GWTC3rate}, and hence the current upper limit on $\fPBH$ may decrease by a factor $\sim 3$~\cite{Wong:2020yig}.
In practice, local constraints rely on being able to distinguish the mass and spin spectra between primordial and astrophysical BBHs~\citep{Franciolini:2021tla,Franciolini:2021xbq,Franciolini:2022iaa}.
Therefore, interpreting these \textit{local} BBHs will still be limited by the model uncertainties of other astrophysical formation channels dominating the late time Universe, which heavily depends on the detailed dynamics of binary formation and stellar evolution.
In contrast, the projected upper limit derived in this work takes advantage of the distinctive ``smoking-gun'' feature of the redshift evolution at high redshifts and is only limited by the actual Pop~III merger rate density in the high redshift.

\begin{figure}[!h]
    \centering
    \subfloat[Phenomenological model]{\includegraphics[width=0.92\columnwidth]{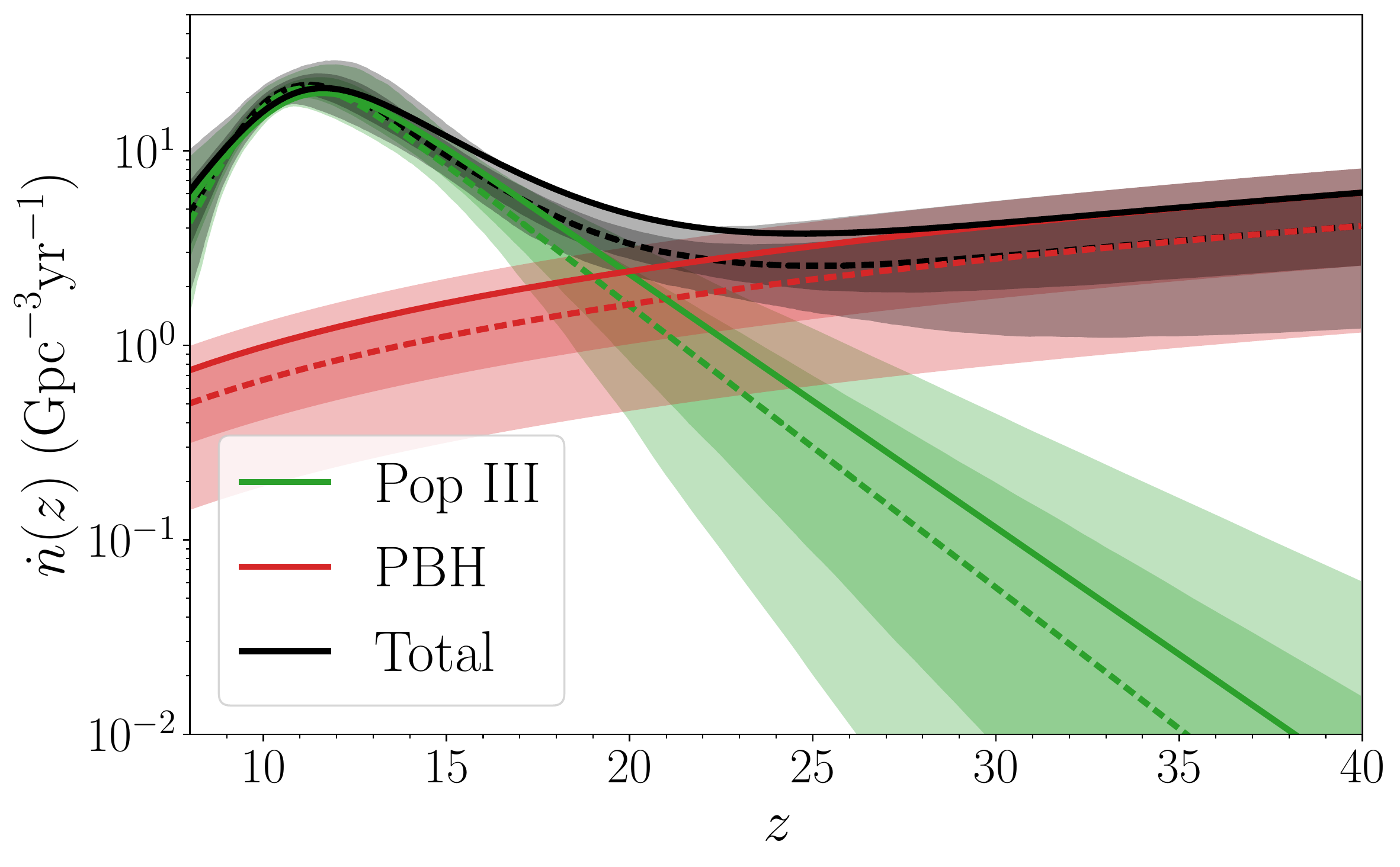}}
    \quad
    \subfloat[Gaussian process]{\includegraphics[width=0.9\columnwidth]{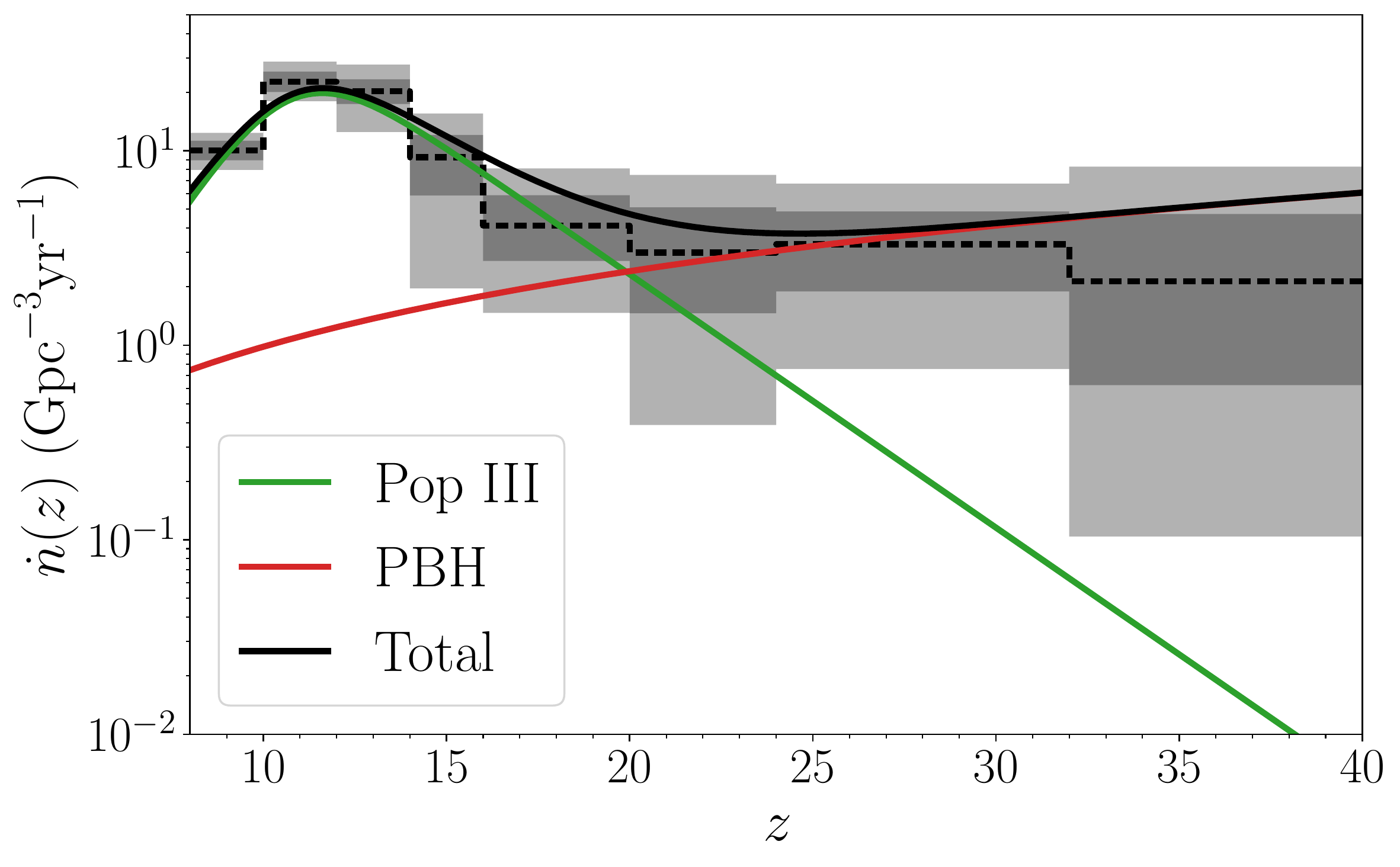}}
    \caption{Posterior of the merger rate densities of the total mergers (black), PBH mergers (green), and Pop~III mergers (red) using phenomenological models (top panel) and Gaussian process regression (bottom panel) in a Universe with $\betaPBH=1/5$ $(\fPBH\approx 10^{-4})$.
    The darker (light) color band corresponds to 68\% (95\%) CI.
    The dashed and solid lines indicate the median of the inferred rate densities and the true rate densities, respectively.
    Since we cannot model each subpopulation in GPR, there are no posteriors of the merger rate densities of the individual subpopulations.
    }
    \label{fig:RateDensitiesPosterior}
\end{figure}

Next, we compare our phenomenological results on the merger rate density -- Fig~\ref{fig:RateDensitiesPosterior} top panel -- with what we obtain using the unmodeled GPR approach (bottom panel) in a universe with $\betaPBH=1/5$ $(\fPBH\approx 10^{-4})$.
With the phenomenological model we can measure the peak of the Pop~III merger rate density within $\sim 30\%$ relative uncertainty, consistent with previous studies~\cite{Ng:2020qpk}.
The 95\% CIs of the merger rate densities of the two populations at $z\gtrsim 30$ do not overlap with each other.
With the GPR model we obtain a similar result, except the posterior for each redshift bin has a larger relative uncertainty, which grows to a factor of 10 in the last redshift bin $32\leq z<40$.
In all cases, the true merger rate densities lie within the 95\% CI.
In Appendix D, we list the hyperpriors used in both models, and show the full hyperposterior of the phenomenological model for a universe with $\betaPBH=1/5$. We stress that the inference of the hyperparameters, and thus of the merger rate, is not affected by the assumption that $M_c$ and $\sigma$ are known.

We can perform a back-of-the-envelope estimate of how increasing the Pop~III merger rate would affect the upper limit on $\fPBH$.
While the noise floor increases with the actual Pop~III merger rate density, the uncertainty of the total merger rate also decreases with more detectable sources.
This uncertainty may be directly translated to an upper limit on $\fPBH$, since the excessive fluctuation may be fully misidentified as PBH mergers at worst.
Let us assume the average Pop~III merger rate density is $\nIII \sim r\RateDensityUnit$ in the high-redshift region $25\leq z \leq 40$.
There are two possible sources of uncertainties on the merger rate: Poisson fluctuation (scaling as $\propto \sqrt{r}$) and the individual redshift measurement  uncertainties.
To be conservative, we calibrate the uncertainty of the total merger rate by the posteriors of GPR merger rate model, that is $\sim 4\RateDensityUnit$ at $\dot{n}\sim 4\RateDensityUnit$ between $25\leq z \leq 40$ (Fig.~\ref{fig:RateDensitiesPosterior}).
The expected uncertainty of the total merger rate density in this redshift range is $\sim 4\sqrt{r/4}\RateDensityUnit$ within four months of observations.
For $\fPBH\lesssim 10^{-3}$, we have $\npbh\propto \fPBH^2$, and hence the upper limit on $\fPBH$ scales as $\sim 10^{-4}(r/4)^{1/4}$ for one year of observations.
As a validation of this estimate, the scaled upper limit is $\sim 4\times 10^{-5}$ for $r\sim 0.1\RateDensityUnit$ in our simulation, consistent with the value inferred by the phenomenological model.
The scaling of $r^{1/4}$ causes a slow increase in the upper limit of $\fPBH$.
In other words, our projected upper limit on $\fPBH$ does not weaken and not exceed $\fPBH\lesssim 10^{-4}$, unless the Pop~III merger rate density in the high-redshift region is $\sim 100$ higher than the current estimate suggested by population synthesis studies (e.g.~\citep{Belczynski:2016ieo,Liu:2020ufc,Hartwig:2016nde,Tanikawa:2021qqi,Hijikawa:2021hrf}).

\section{Discussion}\label{Sec.Dis}
We have shown that a measurement of the redshift evolution of high-redshift BBHs detectable by NG detectors can lead to a two-orders-of-magnitude improvement on the constraints of the PBH abundance relative to the current constraints derived by the LVKC merger rate measurements.
Considering a mixture of stellar-mass BBHs originated from Pop~III stars and PBHs, we simulated four-months worth data at $z\geq8$ with a network of CE in the US and ET in Europe.
Using hierarchical Bayesian inference with a phenomenological model of the merger rate density, we may measure a non-zero PBH abundance as low as $\fPBH\gtrsim 1\times 10^{-4}$ within an uncertainty of $\sim50\%$.
If, instead, the data does not support an excess of BBHs beyond $z\gtrsim 30$, we may obtain an upper limit of $\fPBH\sim\mathcal{O}(10^{-5})$.
The total merger rate density inferred by a Gaussian process model, while showing a larger uncertainty, is also consistent with the phenomenological analysis.

The inference of $\fPBH$ benefits mainly from the distinctive features in the redshift evolution of the Pop~III and PBH merger rates: as the redshift increases, the former decays quickly, while the latter rises monotonically.
This feature helps measuring the relative abundance between the two populations, and hence the PBH abundance from the inferred PBH merger rate density, even in absence of other distinguishing features, e.g. the mass spectra.
Even though the quoted figure of merit depends on the detailed assumptions on the Pop~III merger rate density at high redshifts, our projected upper limit is a conservative estimate for the following reason.
The normalization of the Pop~III rate used in this study already supports a significant number of detectable Pop~III BBHs beyond $z\gtrsim 8$ when compared to Pop~I/II BBHs.
If the actual Pop~III rate were much smaller, the fewer Pop~I/II mergers would become the ``noise floor'', and the derived upper limit of $\fPBH$ would decrease even further.
One may also worry that a PBH population could modify the Pop~III formation scenario studied assuming $\Lambda$CDM.
As shown by recent work~\cite{Liu:2022okz}, the small values of the PBH abundance considered in this work are not expected to have an impact on the formation of Pop~III stars. 

We end by discussing a few caveats associated with our analyses, as well as possible future avenues. 
First, we have assumed that the GWs are dominated by the (2, 2) harmonic, which limits the precision of distance measurements due to distance-inclination degeneracy~\citep{Usman:2018imj,Chen:2018omi}.
Reference~\citep{Ng:2021sqn} shows that including higher-order modes in the waveform modeling may improve the precision of distance measurements by $\sim 30-50\%$ for stellar-mass BBHs at $z\geq 10$, which propagates to an improvement on the measurement of the merger rate density, and hence of $\fPBH$.
Second, we have assumed that the mass and spin spectra between Pop~III and PBH mergers are the same.
This is generally not expected as their formation mechanisms are different in nature.
A joint inference of the distribution of redshift, masses and spins may provide additional information for distinguishing these two high-redshift populations~\citep{Franciolini:2021tla}. 
{This also implies that it is not obvious that measuring $M_c$ and $\sigma$ from the data and marginalizing them over -- instead of fixing them to some value as we have done -- will necessarily make the inference on $\fPBH$ more uncertain. While extra degrees of freedom generally increase the statistical uncertainty in each parameter, information on the mass would provide an additional way of distinguishing the two channels, thus reducing the uncertainty.}

Finally, the main modeling uncertainty comes from the initial conditions of the Pop~III stars, which directly affects the nominal value of the Pop~III merger rate density~\citep{Belczynski:2016ieo,Tanikawa:2021qqi,Hijikawa:2021hrf}.
This may be better constrained by near-future facilities such as the James Webb Space Telescope, Euclid or the Roman space telescope, which may probe the properties of Pop~III stars by gravitational lensing in blind surveys of Pop~III galaxies~\citep{Vikaeus2021}.

\acknowledgements
The authors would like to thank Valerio De Luca, Vishal Baibhav, and Kaze Wong for suggestions and comments.
G.F. and K.K.Y.N. thank Johns Hopkins University for the kind hospitality during the completion of this project.
K.K.Y.N. and S.V., members of the LIGO Laboratory, acknowledge the support of the National Science Foundation through the NSF Grant No.
PHY-1836814. LIGO was constructed by the California Institute of Technology and Massachusetts Institute of Technology with funding from the National Science Foundation and operates under Cooperative Agreement No. PHY-1764464.
E.B. is supported by NSF Grants No. PHY-1912550, AST-2006538, PHY-090003 and PHY-20043, and by NASA ATP Grants No. 17-ATP17-0225 and 19-ATP19-0051. This work has received funding from the European Union’s Horizon 2020 research and innovation programme under the Marie Skłodowska-Curie grant agreement No. 690904.
A.R. is supported by the Swiss National Science Foundation 
(SNSF), project {\sl The Non-Gaussian Universe and Cosmological Symmetries}, project number: 200020-178787.
G.F. and P.P. acknowledge financial support provided under the European Union's H2020 ERC, Starting Grant agreement no.~DarkGRA--757480, under the MIUR PRIN, FARE programmes (GW-NEXT, CUP:~B84I20000100001) and 
H2020-MSCA-RISE-2020 GRU.
This is ET document ET-0068A-22 and CE document CE-P2200005.

\appendix
\section{Hierarchical Bayesian inference with selection effect}\label{app:hba}
Here, we review the mathematical details of the hierarchical Bayesian inference framework~\citep{Farr:2013yna,Mandel:2018mve,Thrane:2018qnx,Wysocki:2018mpo,Vitale:2020aaz}.
In the following, we assume that the mass distribution is non-evolving and focus on the redshift measurements alone.
In both the phenomenological and GPR models, the differential merger rate in the detector frame,
$\dNdz(z\mid \pmb{\Lambda}, R)$,

depends on the hyperparameters $(\pmb{\Lambda}, R)$ that control the morphology and the normalization, respectively.

Given a set of $\Nobs$ observations $\pmb{d}\equiv \{ d_i \}_{i=1}^{\Nobs}$, the posterior of $\pmb{\Lambda}$ in the presence of selection effect is
\begin{align}\label{eq:hyperpos}
& \quad p\left( \pmb{\Lambda}, R \mid \pmb{d}\,\right) \nonumber \\
&\propto \left[\prod_{i=1}^{\Nobs} T \int dz_i \, L_{\rm GW}\left(z_i \right) \dNdz\right]  e^{-R_{\rm det}T} \,\pi\left(\pmb{\Lambda}, R \right) \nonumber \\
&\simeq \left[\prod_{i=1}^{\Nobs} \left( \frac{T}{M_i} \sum_{j=1}^{M_i} \dNdz \left(z_{ij} \mid \pmb{\Lambda}, R \right) \right)\right] \\ \nonumber
&\quad \times e^{-R_{\rm det}(\pmb{\Lambda}, R) T} \, \pi\left(\pmb{\Lambda}, R \right),
\end{align}
where $\left\{z_{ij}\right\}_{j=1}^{M_i}$ is the set of $M_i$ samples drawn from the individual likelihood of the $i$-th redshift measurement $L_{\rm GW}(z_i) \equiv p(d_i \mid z_i)$ obtained from GW observation.
We approximate the integrals in the second line by an importance sum in the third line of Eq.~\eqref{eq:hyperpos}.
The remaining pieces are the hyperprior on the hyperparmeters, $\pi(\pmb{\Lambda}, R)$, the time window over which observations occur in the detector frame, $T$, and the detectable total rate $R_{\rm det}$ that accounts for the selection effect~\cite{Mandel:2018mve,Thrane:2018qnx,Wysocki:2018mpo,Vitale:2020aaz},
\begin{align}\label{eq:Rdet}
    R_{\rm det}(\pmb{\Lambda}, R) = \int \frac{dR}{d\theta}(\theta \mid \pmb{\Lambda}, R) p_{\rm det}(\theta) d\theta,
\end{align}
where $p_{\rm det}(\theta)$ is the detection efficiency of sources with a full set of parameters $\theta$ (including masses, spins, redshift, etcetera).
In our simulation, a source is detectable if its network SNR is larger than 12.
To estimate $R_{\rm det}$, we first perform a Monte Carlo simulation of the sources drawn from a baseline population that can surpass the SNR threshold.
The baseline redshift distribution is constant in merger rate density, $p_0(z)\propto dV_c/dz/(1+z)$.
The mass distribution is same as in the simulations (see Eq.~(\ref{eq:masspsi}) in the main text)
and we conservatively assume that measurements of the mass spectrum are uninformative.
Hence, the selection effect due to the variation of the mass spectrum only changes the total number of observations.
This has been accounted for in the scaling of the projected uncertainties shown in the main text.
We fix the spins to be zero, and use uniform priors on other parameters, such as the sky location, orbital orientation and polarization angle.
Once we have a set of parameters of $N_{\rm sel}$ sources that surpass the SNR threshold, $\{\theta_i\}_{i=1}^{N_{\rm sel}}$, we can evaluate the integral in Eq.~\eqref{eq:Rdet} by importance sampling and reweighing the baseline population prior.
In summary, we approximate Eq.~\eqref{eq:Rdet} in the redshift-only analysis to be 
\begin{align}
    R_{\rm det}( \pmb{\Lambda}, R) \simeq \sum_{i=1}^{N_{\rm sel}} \dNdz(z_i \mid \pmb{\Lambda}, R)\frac{1}{p_0(z_i)}.
\end{align}

\section{Gaussian Process Regression}\label{app:GPR}
This section provides details on the implementation of the GRP model to infer $\dot{n}(z)$ over finite redshift bins.
We define $dR/dz$ as a piecewise-constant function over $W = 8$ redshift bins between $8\leq z \leq 40$.
The bins are uniformly distributed in each redshift range: 4 bins in $[8, 16]$, 2 bins in $[16, 24]$, and 2 bins in [24, 40].
We intentionally choose coarse intervals to make sure each bin has enough samples and avoid numerical fluctuations.
The merger rate $\dNdz$ is thus
\begin{equation}
\frac{dR}{dz}
= \begin{cases}
\frac{\Delta R_1}{\Delta z_1} & 8 \leq z < z_1 \\
\ldots & \\
\frac{\Delta R_i}{\Delta z_i} & z_{i-1} \leq z < z_i \\
\ldots & \\
\frac{\Delta R_W}{\Delta z_W} & z_{W-1} \leq z < 40
\end{cases},
\end{equation}
where $\Delta R_i$ is the merger rate in $i$-th redshift bin $\Delta z_i~{\equiv}~z_i - z_{i-1}$, so that $\sum_{i=1}^W (\Delta R_i)~{\equiv}~R$.
Following Ref.~\citep{Ng:2020qpk}, we infer $\dNdz$ in natural-log space for computational efficiency.
The Gaussian process prior is based on a squared-exponential kernel on $X_i \equiv \ln{(\Delta R_i)}$ that prevents from over-fitting~\citep{Foreman-Mackey:2014},
\begin{align}
K_{ij} &\equiv \mathrm{Cov}\left( X_i, X_j \right) \nonumber \\
&= \sigma_{X}^2 \exp\left[-\frac{1}{2}\left(\frac{z_{i-1/2} - z_{j-1/2}}{l}\right)^2\right],
\end{align}
where $z_{i-1/2} = \frac{1}{2}\left( z_{i} - z_{i-1} \right)$ is the midpoint of the $i$-th redshift bin, $\sigma_X^2$ is the variance of $\{X_i\}$, and $l$ is the correlation length in redshift space.
The multivariate Gaussian process prior on the vector of binned merger rate $\pmb{X} \equiv \{\ln{(\Delta R_i)}\}$ with a mean vector $\pmb{\mu}_X$ and a covariance matrix $\mathbf{K}\equiv \{K_{ij}\}$ is
\begin{align}
\mathcal{G}(\pmb{X} \mid \pmb{\mu}_X, \sigma_X,l) \equiv \mathcal{N}\left[ \pmb{X} \mid \pmb{\mu}_X, \mathbf{K}\left(\sigma_X,l\right) \right],
\end{align}
and we decompose $\mathbf{K}$ into a lower-triangular matrix $\mathbf{L}$,
\begin{align}
\pmb{X} \equiv \pmb{\mu}_X + \mathbf{L}(\sigma_X,l)\pmb{\eta},
\end{align}
where $\pmb{\eta}=\{\eta_i\}_{i=0}^W$ follows from a multivariate standard normal distribution.
Since we know that $dR/dz$ is proportional to $dV_c/dz/(1+z)$, we choose the mean vector as
\begin{align}
\{\mu_{X,i}\} &= \left\{ \ln\left(\frac{\Nobs\Delta V_i}{V}\right) + \Delta\mu_X \right\}, \\
\Delta V_i &= \int_{z_{i-1}}^{z_i}\frac{1}{1+z} \frac{dV_c}{dz} dz, \\
V &= \int_{8}^{40}\frac{1}{1+z} \frac{dV_c}{dz} dz,
\end{align}
so that the variable $\Delta \mu_X$ captures the fluctuation of $R$.

\section{Calibration of the mock likelihood}\label{app:mockLike}
We approximate the redshift likelihood as a lognormal distribution.
Calibrated by the simulations in Ref.~\citep{Vitale:2016icu} using \textsc{IMRPhenomPv2} waveform~\citep{Hannam:2013oca,Husa:2015iqa,Khan:2015jqa}, the relative uncertainty of redshift scales linearly with the true redshift~\citep{Ng:2020qpk}.
This uncertainty relation is only valid for the redshift range with negligible selection effect.
At larger redshift, the SNRs of the sources are close to the threshold, and the redshift uncertainties no longer follow the linear scaling.
In this study, we instead calibrate the standard deviation of the lognormal, $\sigma_{\rm LN}$, as a function of network SNR, $\rho_{\rm net}$:
\begin{align}
    \sigma_{\rm LN} = 1.41 \rho_{\rm net}^{-0.74} \nonumber
\end{align}
for the detector network consisting of CE in the US and ET in Europe.
This returns $\sim20\%$ 1-sigma relative uncertainty for $\rho_{\rm net}=12$.
We note that this fit only reflects the median of the uncertainties in the simulation set~\citep{Vitale:2016icu}.
The actual uncertainty obtained by full Bayesian inference can deviate by a factor of two from the fit at each $\rho_{\rm net}$.
This is because using $\rho_{\rm net}$ alone cannot capture the correlation between redshift and other extrinsic parameters, which may deform the marginalized redshift likelihood.

\section{Hyperpriors and hyperposteriors}
\begin{figure}[t]
    \centering
    \includegraphics[width=0.9\columnwidth]{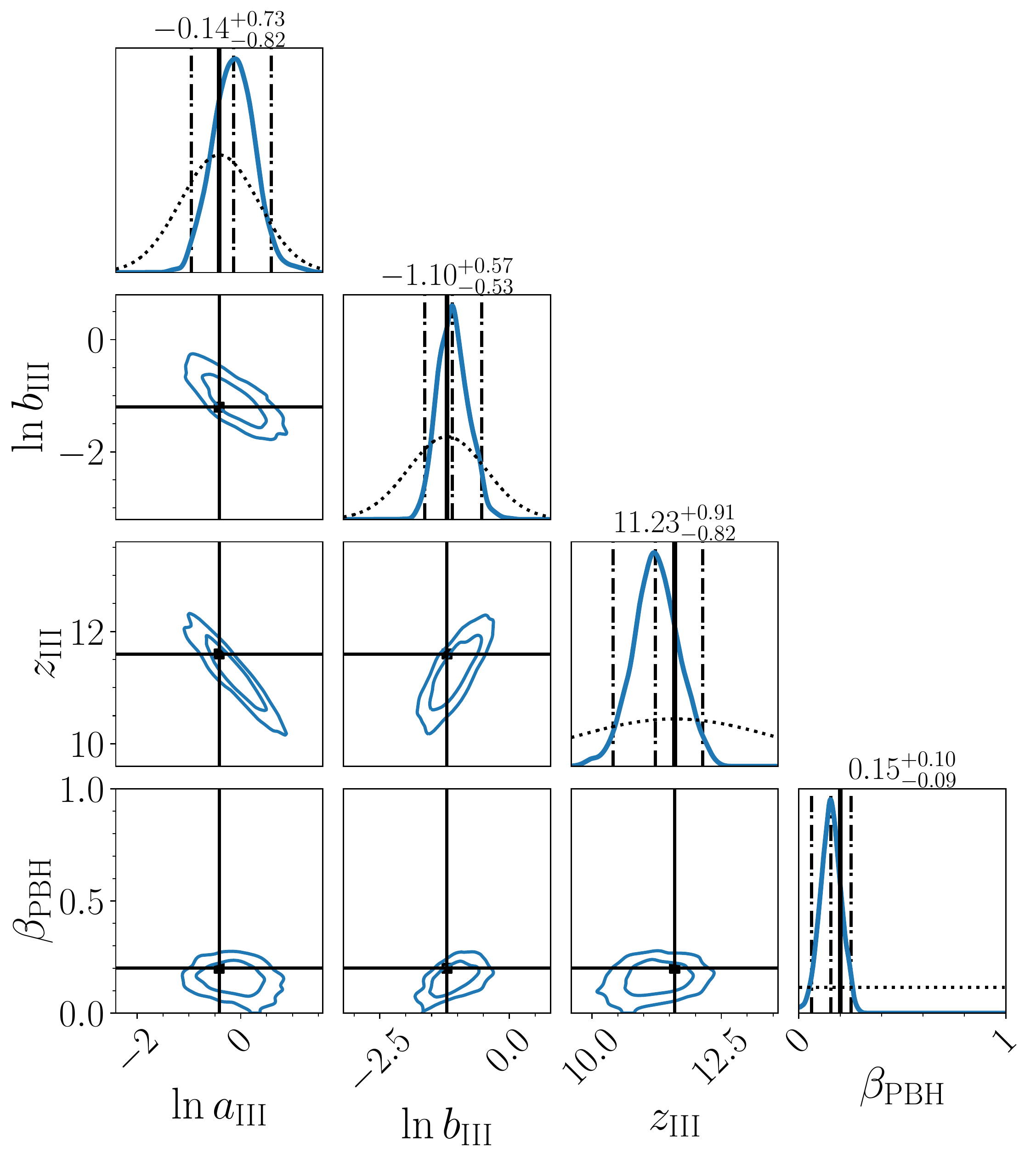}
    \caption{Corner plot of the posteriors of the four hyperparameters, $(\aIII,\bIII,\zIII,\betaPBH)$, for a Universe with $\betaPBH=1/5$.
    2D contours and black dashed lines represent the 68\% and 95\% CIs of the marginalized 2D and 1D posteriors, respectively.
    True values are marked by black solid lines.
    Dotted lines in the diagonal slots show the uncorrelated priors on the hyperparameters.
    }
    \label{fig:PhenomPos}
  \end{figure}

The priors on the hyperparameters of GPR model and phenomenological model, $\Lambda_{\mathrm{GPR},i}$ and $\Lambda_{\mathrm{PM},i}$,  are tabulated in Tables~\ref{tab:GPRpriors} and~\ref{tab:PMpriors}, respectively.
Figure~\ref{fig:PhenomPos} shows the corner plot of the posteriors of the four hyperparameters, $(\aIII,\bIII,\zIII,\betaPBH)$, for a Universe with $\betaPBH=1/5$.

\begin{table}[t]
\begin{ruledtabular}
\caption{Hyper-priors for the GPR model.
}
\label{tab:GPRpriors}
\begin{tabular}{cccc}
$\Lambda_{\mathrm{GPR},i}$ & Prior function & Prior parameters & Domain \\
\hline
$\eta_i$ & Normal & $(\mu_{\mathrm{N}},\sigma_{\mathrm{N}})=(0,1)$\footnote{$\mu_{\mathrm{N}}$ and $\sigma_{\mathrm{N}}$ are the mean and standard deviation of the normal distribution, respectively.} & $(-\infty,+\infty)$\\
$\Delta\mu_X$ & Normal & $(\mu_{\mathrm{N}},\sigma_{\mathrm{N}})=(0,10)$ & $(-\infty,+\infty)$\\
$\sigma_X$ & Lognormal & $(\mu_{\mathrm{LN}},\sigma_{\mathrm{LN}})=(0,4)$ & $(0,+\infty)$\footnote{$\mu_{\mathrm{LN}}$ and $\sigma_{\mathrm{LN}}$ are the mean and standard deviation of the lognormal distribution, respectively.}\\
$l$ & Lognormal & $(\mu_{\mathrm{LN}},\sigma_{\mathrm{LN}})=\left(0,\frac{1}{2}\ln(10) \right)$ & $(0,+\infty)$\\
\end{tabular}
\end{ruledtabular}
\end{table}

\begin{table}[t]
\begin{ruledtabular}
\caption{Hyper-priors for the phenomenological models.
}
\label{tab:PMpriors}
\begin{tabular}{cccc}
$\Lambda_{\mathrm{PM},i}$ & Prior function & Prior parameters & Domain \\
\hline
$\aIII$ & Lognormal & $(\mu_{\mathrm{LN}},\sigma_{\mathrm{LN}})=(\hat{a}_{\rm III},0.5)$ & $(0,2]$\\
$\bIII$ & Lognormal & $(\mu_{\mathrm{LN}},\sigma_{\mathrm{LN}})=(\hat{b}_{\rm III},1)$ & $(0,2]$\\
$\zIII$ & Normal & $(\mu_{\mathrm{N}},\sigma_{\mathrm{N}})=(\hat{z}_{\rm III},2)$ & $[8,20]$\\
$\betaPBH$ & Uniform & --- & $[0,1.0]$\\
$RT$ & Half Cauchy & $\gamma_{C}=\Nobs$\footnote{$\gamma_{C}$ is the scale parameter of the Cauchy distribution.}& $[0,+\infty]$
\end{tabular}
\end{ruledtabular}
\end{table}

\clearpage
\bibliography{pbh.bib}

\end{document}